\documentclass[aps,prb,superscriptaddress,showpacs,floatfix]{revtex4-2}
\usepackage{graphicx,amsmath,amssymb,xspace,epsfig,float,multirow,subfigure,tabularx}
\usepackage{amsbsy,amssymb,amsmath,bm}
\usepackage{epsf}
\usepackage{color,natbib}
\usepackage{amsfonts}
\usepackage{hyperref}

\pdfoutput=1
\input{tcilatex}
\begin{document}

\title{Maximal critical temperature dependence on number of layers due to
phonon d-wave pairing in hole doped cuprates.}
\author{Baruch Rosenstein}
\affiliation{Department of Electrohysics, National Yang Ming Chiao Tung University, Hsinchu,
Taiwan, R.O.C. }
\email{vortexbar@yahoo.com}
\author{B. Ya. Shapiro}
\affiliation{Department of Physics, Institute of Superconductivity, Bar-Ilan
University, 52900 Ramat-Gan, Israel.}
\email{shapib@biu.ac.il}
\author{Guy Leshem}
\affiliation{Department of Computer Science,Academic Colledge of Ashkelon, 78211, Ashkelon, Israel.}
\email{gialsm@edu,aac.ac.il}

\begin{abstract}
Recently an apical oxygen atoms vibrations exchange mechanism of d-wave
pairing in cuprates was proposed. The phonon mode in an insulating layer
generates attraction of \ holes in metallic $CuO_{2}$ plane with potential $%
V\left( k\right) \propto \exp \left[ -2kd_{a}\right] $, where $d_{a}$ is the
distance to the apical plane. The pairing has a maximum at the
crystallographic $\Gamma $ point leading to d-wave channel. The idea is
generalized here to include the $CuO_{2}$ breather and half - breather \
modes in a multi-layer cuprate generating the pairing in an adjacent $CuO_{2}
$ layer of the same multi-layer. It is demonstrated that the phonon exchange
and the spin fluctuation pairing constructively enhance each other since the
paramagnon pairing peaks at crystallographic $M$ point. The phonon
contribution explains the maximal critical temperature $T_{c}^{\max }$
dependence on the number of layers $N$. It rises equidistantly by $15K$ from 
$N=1$ to $N=3$ and then saturates. The strength of the onsite Coulomb on
site repulsion at optimal doping is to obtain the observed values of $%
T_{c}^{\max }$ in the intermediate range of the effective on-site repulsion $%
U=\left( 1.5-2\right) \ eV$, smaller than commonly used in purely in-plane
(spin fluctuation) theory of high $T_{c}$ superconductivity.
\end{abstract}

\pacs{PACS: 74.20.Rp,   74.72.-h,  74.25.Dw}
\maketitle

\affiliation{Department of Electrohysics, National Yang Ming Chiao Tung
University, Hsinchu, 30050, Taiwan, R.O.C. }

\affiliation{Department of Physics, Institute of Superconductivity, Bar-Ilan
University, 52900 Ramat-Gan, Israel.}

\affiliation{Department of Computer Science, Ashkelon Academic College,
78211 Ashkelon, Israel.}

\affiliation{Electrophysics Department, National Yang Ming Chiao Tung 
University, Hsinchu 30050, Taiwan, R. O. C}

\affiliation{Physics Department, Bar-Ilan University, 52900 Ramat-Gan,
Israel}

\section{Introduction}

The physical cause of high temperature superconductivity in cuprates is
still under intensive debate despite enormous body of experimental facts and
numerous theoretical proposals. One of the reasons is the fact that the
normal state physics is very rich. It contains the anti-ferromagnetic Mott
insulator at low doping, the pseudogap and strange metal phases at
intermediate dopings and the Fermi metal at large doping. There is little
doubt that the later is due to "strong coupling" physics within the $CuO_{2}$
plane electron system with the on site electron - electron repulsion $U$
significantly larger than the nearest neighbor hopping energy $t$ and at
least comparable to the band width $W\approx \left( 3-4\right) eV\approx
\left( 6-8\right) t$. The repulsion does not allow the "conventional" s-wave
pairing due to phonon exchange demonstrated in superconductors. Since the
anti-ferromagnetic spin fluctuations (SF) exchange cause the d-wave pairing,
it was conjectured early on that this purely \textit{in - plane} mechanism
play a major role\cite{SF}. It is well known that the mesoscopic \ Hubbard
model at small couplings, $U<<W/2$, the critical temperature at optimal
doping is according to the RG estimate\cite{RG} exponentially small, $%
T_{c}^{\max }$ $\approx W\exp \left[ -\left( N_{0}U\right) ^{-2}\right] $.
Here the density of states is $N_{0}\approx 1/W$. At strong coupling (the $%
t-J$ limit\cite{Plakida}), $U>>W/2$, $T_{c}^{\max }$ decreases slower (still
exponentially) with $U$, $T_{c}^{\max }$ $\approx \sqrt{Wt}\exp \left[
-\left( N_{0}J\right) ^{-1}\right] $, where the exchange energy is $J\approx
4t^{2}/U$. Therefore the maximum occurs at a moderate value around $U\approx
W/2$, see Fig.1. Whether the SF d - wave pairing \textit{alone} can reach
the value of $T_{c}^{\max }$ above $160K$ observed in best multi-layered
materials like $HgBa_{2}Ca_{2}Cu_{3}O_{8}$ under pressure is still and open
question. In addition there is an evidence\cite{Kao23} that the observed the
pseudogap physics and the strange metal resistivity (firmly believed to be
caused by the SF) can be described by values of $U$ \textit{below} $W/2$.

The SF is not the only possibility to induce d-wave pairing in these
systems. Recently it was pointed out\cite{I} within a phenomenological Born
- Mayer type model\cite{Falter} that an optical phonon due to the lateral
apical oxygen vibration (the AP mode) couples to the $CuO_{2}$ plane
electrons in such a way that it also induces the d-wave pairing. The d-wave
component reflects a so called "central peak" in forward scattering \ (a
possibility considered \cite{Kulic} long ago) due to spatial separation of $%
d_{a}=2.7A$ between $CuO_{2}$ and the adjacent insulating oxide layers
containing the apical oxygen ions, $V\left( q\right) \propto \exp \left[
-2qd_{a}\right] $. The atomic () charges required within this approach were
taken from first principle (LDA DFT) calculations\cite{Evarestov}.

If such a pairing is present and works together with the SF exchange, the
value of $U$ can be smaller than "optimal" one, $W/2$. At relatively small
couplings notoriously difficult interactions effects requiring Monte Carlo
or tensor products\cite{MC} methods can be treated by simpler methods. In
the framework the renormalized perturbation theory\cite{Schrieffer} adding
the SF exchange at relatively small $U<W/2$, can give\cite{I} $T_{c}^{\max
}\approx 90K$ for parameters of $Bi_{2}Sr_{2}CaCu_{2}O_{8}$. The
constructive pairing impact of the two "bosons" is due to the separation in
momentum of the phonon (central peak pairs, the $\Gamma $ point) from the SF
active near the $M$ point. Note that the d-wave pairing is sensitive to the 
\textit{gradient} of the pairing potential (only the s-wave potential is
sensitive the momentum independent part, but this is effectively quenched in
cuprates even for repulsion as small as $U=2t$).

Meantime more advanced \textit{ab initio} calculations of the electron -
phonon matrix elements (EPC) $g(\mathbf{k,q})$ for the states on the Fermi
surface appeared\cite{Laoie21}\cite{Marzari23}. The GW improvement generally
gives larger values of EPC on Fermi surface than the traditional LDA\cite%
{Laoie08}. It was found that in cuprates there generally three modes with
EPC significantly larger others. These are the $CuO_{2}$ planar high energy, 
$\Omega _{p}\approx \left( 70-80\right) meV$, longitudinal optical modes and
the apical oxygen vibrations (AP) at a lower frequency $\Omega _{a}\approx
55meV$ considered in ref. \cite{I}. The two surface modes, the full
breathing mode (FB), four oxygen atoms vibration around the heavier $Cu$
atom) and half breathing mode (HB) only two four oxygen atoms vibrate) were
studied. The values of EPC range $g=70-400meV$ and depend weakly on doping.
The larger values are quite typical to perovskites as was convincingly
demonstrated in ref.\cite{Laoie19} by a systematic study of simpler
materials. These values are consistent with known experimental probes of EPC
like ARPES \cite{Zhou23} or RIXS \cite{Huang21}.

In the present paper it is pointed out that in a multi-layered cuprate the
two $CuO_{2}$ phonon modes also contribute to the d-wave pairing in the 
\textit{neighboring} $CuO_{2}$ plane of the multi-layer. The separation is
larder than the apical one, $d_{a}=3.4A$, but the EPC is typically also
larger. We therefore generalize and extend the phonon d-wave pairing theory
(with a contribution of \ SF at intermediate coupling) in three aspects.
First the general minimal model of hole doped cuprate with any number of
layers $N$ is considered. Second a more comprehensive description of phonons
and their electron - phonon coupling is used,namely in addition to the AP
modes, the paring due to FB and HB modes in adjacent conducting layers of a
multi-layer are included. The third improvement concerns the treatment of
SF. Instead of a simplistic intermediate coupling renormalized mean field
method, we utilize a recently proposed\cite{Tremblay23} two particle self
consistent (TPSC+) method. It is consistent with recent Monte Carlo
simulations \cite{MCHub22} and is expected to be more accurate in a wider
range of couplings and temperatures.

The generalization allows us to explain the following curios property of the
hole doped cuprate superconductors\cite{Kivelson04}. The critical
temperature at optimal doping $p_{opt}$ (that is quite universal for all the
hole doped materials, $p_{opt}\approx 0.16$) has a characteristic
non-monotonic dependence on the number of layers $N$ in all cuprate
families. It rises from $N=1$ to $N=3$ and then very slowly decreases
saturating at about $N=6$. For example the first 7 members of the highest $%
T_{c}^{\max }$ $Hg$ family $HgBa_{2}Ca_{N-1}Cu_{N}O_{2+2N}$ have the
following critical temperatures: $T_{c}^{\max
}=\{95,110,134,130,126,125,125,...\}K$. Note that the first three values are 
\textit{equidistant}. The $T_{c}^{\max }$\ dependence on $N$ within the
purely SF theory of the multi - layer superconductivity should rely on inter
- layer hopping. Although there were phenomenological (Ginzburg - Landau
type) estimates of the effect\cite{Chacravarty04}, a detailed microscopic
theory has been developed only recently\cite{multilayerth}. The results were
rather disappointing. The relatively large temperature differences are not
reproduced and even tendencies are sometimes opposite to those observed.

The overall effect of the AP and FB/HB modes phonon pairing explains this
peculiar equidistant dependence of superconductivity on the number of layers 
$N=1,2,3$. The model is universal in the sense that small differences in the
dispersion relation at the optimal doping, values of frequencies and EPC of
the phonon modes in various materials will be described by the same values
mentioned above \cite{multilayers}. The only parameters that distinguish
between the families of cuprates therefore will be effective Coulomb
strength $U$ and the sequence of layers (including the number of layers $N$%
). The approach is generally distinct \ from the purely SF theory in the
sense that values of $U$ are in the intermediate coupling range rather than
in the strong coupling regime due to a significant phonon contribution to
the d-wave pairing.

The rest of the paper is organized as follows. In Section II a minimal
multilayer cuprate model combining an effective one band $t-t^{\prime }$
Hubbard model of the correlated electron gas and the Frochlich approximation
for harmonic relevant optical phonons is presented. The phonons in
multilayer materials and their electron - phonon coupling relevant for the
d-wave pairing are described. In Section III effective electron - electron
interaction leading to pairing in the d-wave channel including both the
phonons and SF in the range of $U<4t$ is derived in the framework of the
dynamic approach. This is utilized in Section IV to construct a BCS - like
superconductivity theory that is applied to a concrete cuprate sequence
describing a $Hg$ family (with highest $T_{c}$). the framework of the
dynamic approach). While in ref. \cite{I} doping dependence was studied, in
the present work we limit ourself to the optimal doping for a layer with
largest superconducting gap (so called outer plane, OP), while for a
tri-layer material \ the critical temperature for internal layer (IP) is
also calculated. In the last Section V results are summarized and briefly
discussed.

\begin{figure}[h]
\centering\includegraphics[width=8cm]{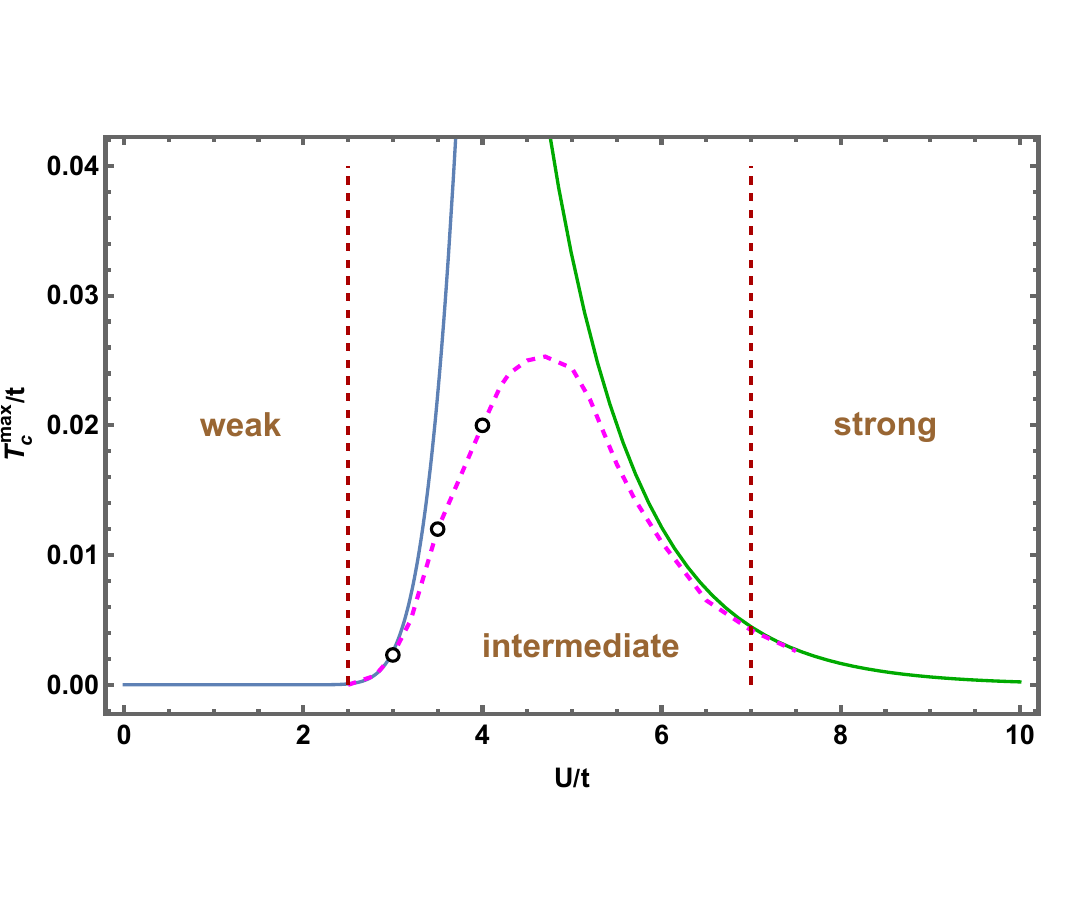}
\caption{ Critical temperature in the Hubbard model as function of the on
site Coulomb repullsion $U$. Both are given in units of the hoping energy $t$%
. Dashed pink line interpolates between the RG weak coupling (blue) and the
t-J strong coupling (green) estimates. Empty circles are the TPSC+ results. }
\label{fig1,pdf}
\end{figure}

\section{The "universal" minimal model.}

Our model consists of $N$\ copies of interacting 2DEG residing in $CuO_{2}$
planes coupled to three kinds of dominant phonon modes residing both inside
and also outside the planes. The structure of the compounds is assumed to be
quite similar with fourfold symmetry and the lattice spacing is almost the
same, $a=3.8A$ (anisotropic in the a-b plane cuprates like $YBCO$ are thus
not considered in the present paper). The basic microscopic geometry of the
crystalline lattice is given in Fig.2, where a single layer cuprate is
shown. The mesoscopic level effective lattice Hamiltonian therefore has a
form:%
\begin{equation}
H=H_{e}+H_{ph}+H_{e-ph}\text{.}  \label{Hamiltoniandef}
\end{equation}%
The electron part $H_{e}$ is an extended Hubbard model, phonons $H_{ph}$ are
described within the harmonic approximation, while the coupling between the
electronic and vibrational degrees of freedom, $H_{e-ph}$, is described
within the Frochlich approximation. The three terms are described in turn.

\begin{figure}[h]
\centering\includegraphics[width=8cm]{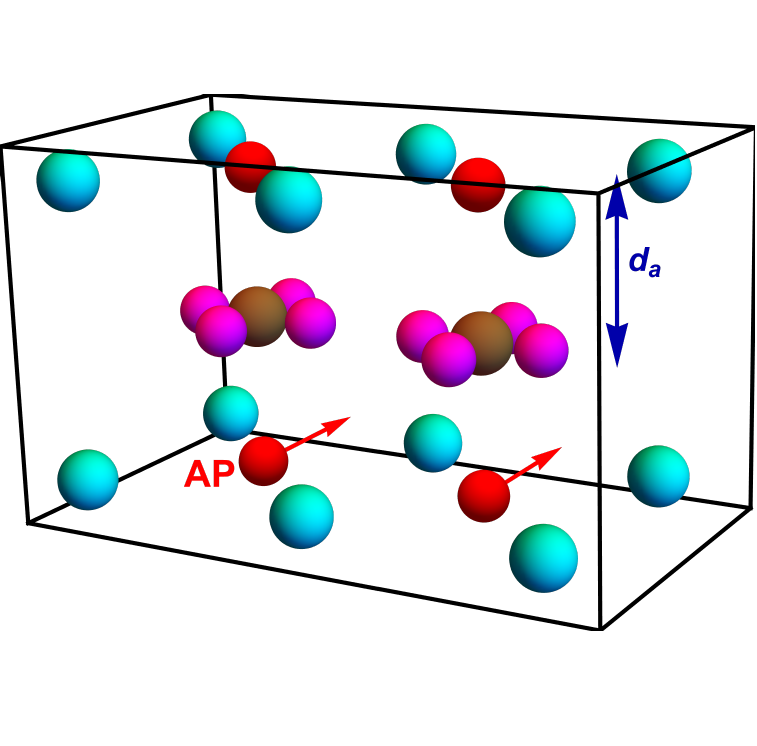}
\caption{Schematics picture of the relevant part of a single layer cuprate $%
HgBa_{2}Cu_{2}O_{4}$. Metallic atoms are: $Cu$ - brown, $Ba$ - magenta$,Hg$-
cyan. Oxygen atoms (small spheres): in the $CuO_{2}$ plane - red, apical $%
BaO $ planes - pink. Red arrows indicate the lateral apical phonon mode (AP)
polarization. }
\label{fig2.pdf}
\end{figure}

\subsection{The minimal ($t$,$t^{\prime }$,$U$) extended Hubbard model for
the electron system in $CuO$ layers.}

In modern condensed matter theory a model is generally constructed via two
step approach. First a "first principle" (density functional based)
calculation tries to establish a sufficiently accurate "mesoscopic"
effective model from which physical observables are derived using great
variety of analytical and numerical methods. In the framework of SF theory
the \textit{minimal} model contains three essential energies on the square
lattice describing the $CuO_{2}$ planes of cuprates, hopping energies $%
t,t^{\prime }$\ and the\ site repulsion $U$. Of course this mesoscopic
theory, a lattice theory based on selection of "relevant" orbitals is not
precise since many hopping (one body) and interaction (two or mode body)
terms are left out. Derivation of such a minimal model from the microscopic
Hamiltonian (using some variants of the DFT approach) is described in ref.%
\cite{DFTmin}.

The minimal description of the electron gas in the $N$-layer material
therefore consists of $N$ copies of the one-band $t-t^{\prime }$ Hubbard
Hamiltonian:

\begin{equation}
H_{e}=K+\frac{U}{2}\tsum\nolimits_{\mathbf{i,}l}n_{\mathbf{i}}^{l}n_{\mathbf{%
i}}^{l}\text{.}  \label{Hamiltonian}
\end{equation}%
Here the $2D$ lateral position is specified by integers $\mathbf{i=}\left\{
i_{x},i_{y}\right\} $, the index $l=1,...N$, labels $CuO_{2}$ planes in a
multi - layer. The density (in units of lattice spacing $a$ used throughout
the paper) is $n_{\mathbf{i}}^{l}=\tsum\nolimits_{\sigma }c_{\mathbf{i}%
}^{l\sigma \dagger }c_{\mathbf{i}}^{l\sigma }$, with $c_{\mathbf{i}%
}^{j\sigma \dagger }$ being the electron creation operator for the spin
projection $\sigma =\uparrow ,\downarrow $. The last term in Eq.(\ref%
{Hamiltonian}) describes the on site repulsion.

\begin{figure}[h]
\centering\includegraphics[width=8cm]{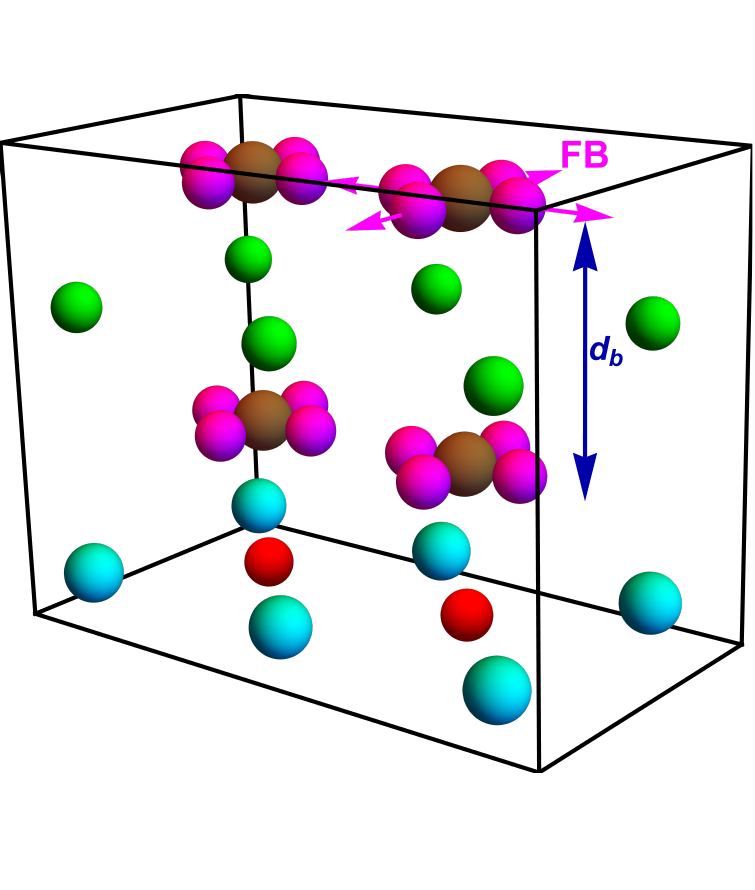}
\caption{A bi-layer cuprate atomic composition. Notations are the same as in
Fig. 2. Polarization of \ the lateral $CuO_{2}$ plane full breazer phonon
mode (FB) is depicted by red arrows. Distance to a metallic layer in which
the d-wave pairing is generates is marked}
\label{fig3.pdf}
\end{figure}

The first limitation of the minimal model pertains to the one body term $K$:
only nearest and next to nearest \textit{in-plane} neighbors hopping terms
are taken into account in $K$, see Fig.2. The dispersion relation thus is
simplified with respect to a "realistic" one in which more distant hops are
included. In momentum space the in-plane hoping Hamiltonian therefore reads: 
\begin{eqnarray}
K &=&\sum\nolimits_{\mathbf{i}\sigma s}c_{\mathbf{k}}^{l\sigma \dagger
}\left( \epsilon _{\mathbf{k}}-\mu ^{l}\right) c_{\mathbf{k}}^{l\sigma };
\label{eps} \\
\epsilon _{\mathbf{k}} &=&-2t\left( \cos \left[ ak_{x}\right] +\cos \left[
ak_{y}\right] \right) -4t^{\prime }\cos \left[ ak_{x}\right] \cos \left[
ak_{y}\right] \text{.}  \notag
\end{eqnarray}%
Moreover there are hops between the layers of a multi-layer. For example in
a bi-layer material, see Fig.3, leading out off - plane hopping term\cite%
{Bansil05} generally has a form CCC $t_{\perp }\left( \cos ak_{x}-\cos
ak_{y}\right) ^{2}$. In our case since the amplitude $t_{\perp }<<t,$ it can
be effectively absorbed as a small correction to the $t^{\prime }$ term. CCC
Therefore the minimal model neglects small splitting in the dispersion
relation seen in ARPES experiments.

On the other hand the "layer chemical potential" might be different $\mu
^{\prime }$. It describes an inequivalent doping\cite{Kondo} for $N>2$. For
example, the doping on inequivalent layers the OP (outer planes) and on IP
(inner plane) of the tri-layer material $HgBa_{2}Ca_{2}Cu_{3}O_{8}$ is
different\cite{Wen24}, see Fig.4.

\begin{figure}[h]
\centering\includegraphics[width=8cm]{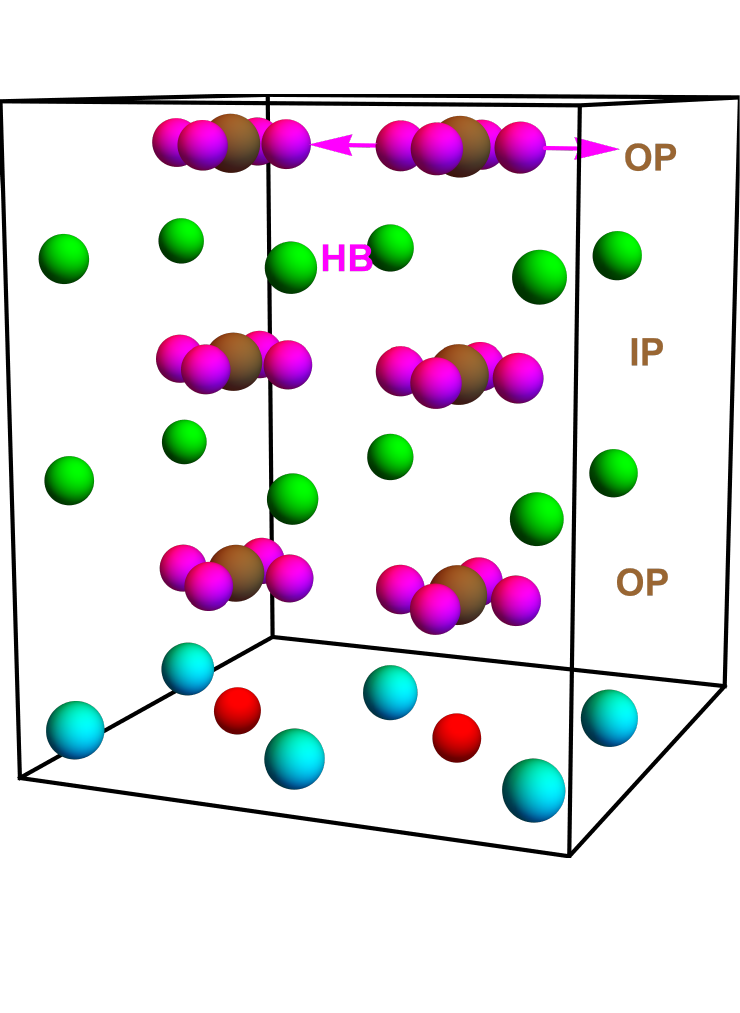}
\caption{ Tri-layer cuprate. Notations are the same as in Fig. 2.
Polarization of \ the lateral $CuO_{2}$ plane half breazer phonon mode (FB)
is depicted by red arrows. }
\label{fig4.pdf}
\end{figure}

Measured values of the lattice in-plane lattice spacing $a=\left(
3.8-3.9\right) A$ is quite universal and will be used as unit of length in
what follows. It turns out that parameters $t$ and $t^{\prime }\ $are also
rather universal in all the hole doped materials. As shown by \textit{ab
initio} determinations for both the minimal\cite{DFTsmallU,DFTmin} and
extended models \cite{DFText,DFTlargeU} the hopping energy to the nearest
site $t$ vary in a limited range $t\approx (360-490)meV$, while the next to
nearest neighbors hopping $t^{\prime }=-\left( 0.15-0.19\right) t$, in
practically all materials. In calculations we use the values of parameters
for the $Hg$ family obtained in the first principles calculation in the
effective theory is reduced precisely to the $t-t^{\prime }$ model with
single interaction term\cite{DFTsmallU}, $t=480meV$, $t^{\prime }=-$ $0.17t$.

The second limitation of the minimal model is that one effectively accounts
for longer range density-density two body terms, other two-body
interactions, and higher interaction terms effects via effective on site
repulsion $U$ only. Its value is not universal and is discussed next
discussed below. In section V we briefly discuss models beyond minimal
typically derived in the framework of DFT with screening, see for example
refs.\cite{DFText,DFTlargeU}, that contain more terms and can in principle
influence the d-wave pairing.

\subsection{ Values of the effective on site repulsion $U$.}

Recent first principle calculations of cuprates provide an increasingly
accurate representation of the effective Hubbard - like model on the
mesoscopic scale utilized in the present paper. Most of them calculate the
parameters for \textit{pristine }materials, namely for zero doping Mott
insulator. Most of the recent determinations \cite{DFText,DFTlargeU} of the
effective Hamiltonian include terms beyond this model for both the hopings
and interactions. We utilize however ref. \cite{DFTsmallU} for multi - layer 
$HgBa_{2}Ca_{N-1}Cu_{N}O_{2+2N}$ most suitable for the present work since we
limit ourselves to the minimal Hubbard model. Values of the on site
repulsion at zero doping of this work are given in Table I as $U_{0}$.

\begin{table*}[h]
\caption{Material parameters of one layer cuprates.}
\begin{center}
\begin{tabular}{|l|l|l|l|}
\hline
material & $T_{c}^{\max }$ & $U_{0}/t[Jang]etal$ & $U/t$ \\ \hline
$HgBa_{2}CuO_{4+\delta }$ & $95$ & $4.52$ & $3.62$ \\ \hline
$HgBa_{2}Ca_{2}Cu_{2}O_{6+\delta }$ & $120$ & $4.49$ & 3.59 \\ \hline
$HgBa_{2}Ca_{3}Cu_{3}O_{8+\delta }(IP)$ & $134$ & $2.41$ & 1.92 \\ \hline
$HgBa_{2}Ca_{3}Cu_{3}O_{8+\delta }(OP)$ & $99$ & $3.07$ & 1.96 \\ \hline
\end{tabular}%
\end{center}
\end{table*}

The minimal model fitting for the $Hg$ sequence ranges from $U_{0}\approx
4.5t$ for the one-layer $HgBa_{2}CuO_{4}$ and two-layer $%
HgBa_{2}CaCu_{2}O_{6}$ to $U_{0}\approx 3.5t$ for the three-layer material $%
HgBa_{2}Ca_{2}Cu_{3}O_{8}$. Generally fitting the DFT with a detailed
mesoscopic model\cite{DFTlargeU} give larger values of $U_{0}$, but ratios
of $U_{0}$ between various materials are practically the same. The later is
useful in that within the non minimal model fitting, doped material
calculations appeared recently\cite{DFText}. It demonstrates that a the
optimal doping, $p_{opt}=0.16$, the value of $U_{opt}$ in the one - band
model is smaller than for zero doping by about 20\% compared to $U_{0}$.
This is also shown in Table I as $U$. As a result the optimal doping
coupling ranges in the lower part of the "intermediate coupling", see Fig.
1, $U=\left( 3-4\right) t$. This range although far from the weak coupling
can be tackled with TPSC+ method\cite{Tremblay23}. We take $U=3t$ to fit the 
$Hg$ material, however consider also $U=3.5t$ and $U=4t$ for which the
method is still applicable for the low temperatures considered (perhaps
approaching the range of applicability). There are other competing
approximation schemes of similar complexity like the covariant \ GW scheme%
\cite{GW} which are not used here. It should be noted that values of $U_{0}$
are dependent on the treatment of screening. Different RPA type methods to
avoiding double counting (model mapped RPA, mRPA, method of ref.\cite{DFTmin}
vs constrained RPA, cRPA, of ref.\cite{DFTsmallU}) influences the result.

\subsection{The three phonon modes with the largest electron-phonon coupling}

Although the prevailing hypothesis is that superconductivity in cuprate is
"unconventional", namely not to be phonon-mediated, the phonon based
mechanism has always been a natural option to explain extraordinary
superconductivity in cuprates. The crystal has very rich spectrum of phonon
modes. However very few have a strong coupling to 2DEG and even fewer can
generate lateral (in-plane) forces causing pairing. The most studied phonon
"glue" modes has been the oxygen vibrations within the $CuO_{2}$ plane\cite%
{Bulut}. It was argued\cite{Rosen19} (in the context of high $T_{c}$ one
unit cell $FeSe$ on perovskite substrates where this was observed
experimentally\cite{Guo}) that lateral vibrations of the oxygen atoms in the
adjacent ionic perovskite layer can couple strongly to 2DEG residing in the $%
CuO_{2}$ plane. Note that the mode is different from the apical oxygen $z$%
axis vibrations (which might be essentially anharmonic) much studied early on%
\cite{apicz}.

The phonon Hamiltonian in harmonic approximation is:

\begin{equation}
H_{ph}=\sum\nolimits_{\mathbf{\nu ,q}}\Omega _{\mathbf{q}}^{\nu }b_{\mathbf{q%
}}^{\mathbf{\nu }\dagger }b_{\mathbf{q}}^{\mathbf{\nu }}\text{.}  \label{Hph}
\end{equation}%
Here $b_{\mathbf{q}}^{\mathbf{\nu }}$ is the annihilation operator of the
phonon mode $\nu $. There are two longitudinal optical modes that are nearly
dispersionless corresponding to lateral vibrations of $CuO_{2}$ in-plane
oxygen atoms (by far the lightest atoms in the system), are in the range of $%
\Omega ^{\nu }\sim $ $70-90meV$ called full breezier and half breezier. They
depend weakly on material. Polarizations of the FB and HB are shown by
arrows on Fig. 3 and Fig. 4 respectively. The frequency value of \ $\Omega
^{FB}=80meV$ is used in calculations of the Section IV. The apical oxygen
atom lateral longitudinal vibration in the range of $(45-60)meV$ is shown in
Fig. 2 and was studied in ref. \cite{I}. The frequency value of \ $\Omega
^{FB}=55meV$ is used in calculations of the Section IV.

Now we turn to the electron - phonon interaction. The two apical oxygen
atoms are above and below the $Cu$ atoms. These distances are $d_{a}=2.4A$
and for the lower $T_{c}$ materials ($La$, $Bi$) and$\ d_{a}=2.7A$ for
higher $T_{c}$ materials ($Hg$,$Tl$). We therefore take the later for
calculations of the Section IV. Similarly the distance between layers $%
d_{p}=3.2A$ is used to describe the central peak due to the layer
separation. The Frochlich Hamiltonian reads,

\begin{equation}
H_{eph}=\sum\nolimits_{\mathbf{kq}\nu \sigma l}c_{\mathbf{k}}^{\sigma
l\dagger }c_{\mathbf{k}-\mathbf{q}}^{\sigma l}g_{\mathbf{kq}}^{\nu l}\left(
b_{\mathbf{q}}^{\nu }+b_{\mathbf{q}}^{\dagger \nu }\right) \text{,}
\label{H_EPI}
\end{equation}%
where $g_{\mathbf{kq}}^{\nu l}$ are the EPC matrix elements for electrons in
the layer $l$ (with $z$ coordinate $z_{l}$) for the phonon mode $\nu =$ FB,
HB, AP. In our case we consider three dominant modes associated either with
a $CuO_{2}$ plane (FB, HB) or apical (AP) that also is characterized by the
its $z$ coordinate $z_{\nu }$.

It was approximated in ref.\cite{I} within the Born-Mayer model as%
\begin{equation}
g_{\mathbf{k,q}}^{\nu l}\approx \frac{2\pi Ze^{2}}{a^{2}}\sqrt{\frac{\hbar }{%
M\Omega _{\mathbf{\nu }}}}\rho _{\mathbf{q}}^{d}.  \label{g}
\end{equation}%
Here the charge $Z$ of the oxygen ion of mass $M$ was taken from \textit{ab
initio} calculation\cite{Evarestov}. The form factor $\rho _{\mathbf{q}}$
for apical phonon model was exponential in the ($z$-axis) distance between
the $CuO_{2}$ layer $l$ and the location of vibrating oxygen atoms, $%
d=z_{l}-z_{\nu }$, 
\begin{equation}
\rho _{\mathbf{q,}l}=e^{-2d\left\vert q\right\vert }\text{.}  \label{rho}
\end{equation}%
The exponential decrease is crucial for creating the d-wave pairing, see
ref. \cite{I} for detailed discussion. It is generalized here to the in
plane lateral longitudinal phonon modes. There is no essential difference in
this formula for the planar phonon modes coupling to the electron gas in the
neighboring $CuO_{2}$ layer of each multi-layer. The only difference is the
momentum independent prefactor (determined in DFT calculations by the zero
value momentum transfer $\mathbf{q}=\mathbf{0}$ and different separation of
the phonon and the electron gas layers.

The (relatively strong) EPC of the full and half breathing modes to the
electron gas in the same layer leads to numerous effects (like the "kink" in
dispersion relation\cite{Laoie21} or resistivity\cite{Kao23}), but cannot
result in the d-wave pairing. It is well established that phonons cause
s-wave pairing in low $T_{c}$ materials, however d-wave pairing is possible
when forward scattering peak of Eq.(\ref{rho}) is present. Early work in
this direction was summarized in ref. \cite{Kulic}. The transition
temperature $T_{c}$ as function of the hole doping within the present model
was studied in ref. \cite{I} demonstrates the maximum at $p_{opt}$. We start
from the derivation of the phonon exchange d-wave "potential" (appearing
mainly near the $\Gamma $ point of the phonon BZ).

The electron phonon couplings, $g_{a}$ and $g_{p}$ will be taken in the
range $70-300meV$ according to available \textit{ab initio} simulations and
some experiments (see Introduction). \ Moreover it turns out that at least
near the $\Gamma $ point the lateral oxygen vibrations in the $CuO_{2}$
layer are stronger than in the apical plane $g_{p}/g_{a}\approx 1.3$. In
calculations we keep this ratio and largely consider the value of $g_{a}$ as
a free parameter along with $U$. The difference between the phonons in the
coupling strength is nearly balanced by the difference in the separation
between the vibrating oxygen atoms (see Figs 3,4) and the location of the
electron gas that as explained above determines the scattering central peak
shape. In Section IV we will use a value of $d_{a}=2.7A$ for the apical
plane to an adjacent $CuO_{2}$ layer that is \ smaller than the inter-layer
distance within multi-layers $d_{p}=3.2A$. Of course this ignores small
variations that exist between different materials that makes the approach
less precise, but allows to capture sufficiently well general features of
the phenomena.

A possibility of the inter-layer phonon generated pairing due tunneling of
electrons between metallic layers was studied long ago\cite{Remova}. It was
shown that this type of the electron - phonon coupling does not lead to the
d - wave pairing. Having thus completed the mesoscopic Hamiltonian
description, we now turn to describe the solution method for this
Hamiltonian based on two steps. First the effective electron-electron
interaction via both the SF and the phonon exchange is computed. Then the
Gorkov equations determine the critical temperature of the transition to
superconductor.

\section{Electron - electron effective interactions}

The Matsubara action describing the electron - electron interaction (that
will be taken into account in the gap equation of the next section) in the
minimal model is: 
\begin{equation}
\mathcal{A}_{ee}^{eff}=\frac{1}{2T}\sum\nolimits_{n\mathbf{p}}V_{n\mathbf{p}%
}^{eff}n_{n\mathbf{p}}n_{\mathbf{-}n\mathbf{,-p}}\text{.}  \label{action}
\end{equation}%
The effective electron potential within the minimal model (neglecting
nonlocal interactions) includes three contributions: 
\begin{equation}
V_{n\mathbf{p}}^{eff}=V_{n\mathbf{p}}^{ph}+V_{\mathbf{p}}^{C}+V_{n\mathbf{p}%
}^{corr}\text{.}  \label{Veff}
\end{equation}%
Here the phonon exchange interaction, $V_{n\mathbf{p}}^{ph}$, will take into
account several modes with largest EPC. It is dynamic, namely strongly
depends on frequency $n$. The (screened) direct Coulomb repulsion $V_{%
\mathbf{p}}^{C}$ is relatively strong compared to that in conventional
superconductors and destroys effectively the s-channel pairing. Since it
does not affect the d-wave channel, it will not be important to our
discussion. Finally the (para) magnon exchange potential $V_{n\mathbf{p}%
}^{corr}$ is weakly dependent on the Matsubara frequency $\omega _{n}=\pi
T\left( 2n+1\right) $ and is specified in the last subsection of this
Section. We start with phonon mediated interaction.

\subsection{The phonon contribution to the d-wave pairing}

Since the tunneling is neglected, in multilayer systems the equation
separate into set of independent equations for IP and OP. Let us start with
the one layer materials, see Fig.2, for which one has \textit{two} AP phonon
mediated pairing.

\begin{equation}
V_{n\mathbf{p}}^{ph}=-2g_{a}^{2}e^{-2pd_{a}}\frac{\Omega _{a}}{\omega
_{n}^{b2}+\Omega _{a}^{2}}\text{.}  \label{1L}
\end{equation}%
where $\omega _{n}^{b}=2\pi Tn$. For $N\geqslant 2$ outer $CuO_{2}$ planes,
OP, layers (the only ones for two layer materials), one has two in-plane
phonons and one apical, see Figs. 3,4:%
\begin{equation}
V_{n\mathbf{p}}^{ph}=-\frac{g_{a}^{2}\Omega _{a}}{\omega _{n}^{b2}+\Omega
_{a}^{2}}e^{-2pd_{a}}-2\frac{g_{p}^{2}\Omega _{p}}{\omega _{n}^{b2}+\Omega
_{p}^{2}}e^{-2pd_{p}}\text{.}  \label{2L}
\end{equation}%
This last is presented as an example in Fig. 5 for \ $g_{p}=300meV$. It
exhibits the central peak as in the case of AP presented in ref. \cite{I}.

Finally for the inner planes, IP, for $N\geqslant 3$, only in-plane phonons
contribute:%
\begin{equation}
V_{n\mathbf{p}}^{ph}=-4\frac{g_{p}^{2}\Omega _{p}}{\omega _{n}^{b2}+\Omega
_{p}^{2}}e^{-2pd_{p}}\text{.}  \label{3L}
\end{equation}%
Now we turn to the terms originating in Coulomb interactions within the $%
CuO_{2}$ planes.

\begin{figure}[h]
\centering\includegraphics[width=8cm]{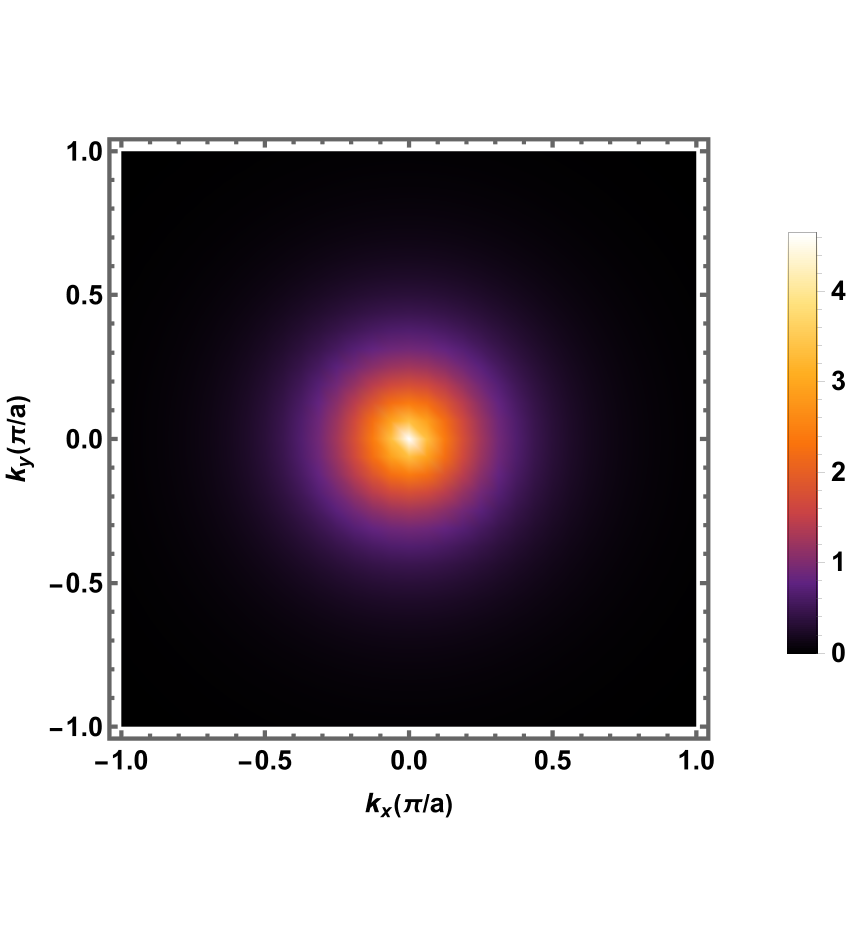}
\caption{ Phonon exchange generated effective electron - electron
interaction. It exhibitits the central peak at crystallographic $\Gamma $
point.}
\label{fig5.pdf}
\end{figure}
As it was noticed above here we use $d_{a}=2.7$ $A$ and $d_{p}=3.2A$ while $%
g_{p}$ and $g_{a}$ will be taken in the range $200-300meV$ and $70-230meV$
correspondingly.

\subsection{ The SF contribution to pairing}

Here we briefly describe a recently developed in ref. \cite{Tremblay23} "two
particles self consistent plus" (TPSC+) approach to Hubbard model in the
intermediate coupling regime $U/t<4$. It considerably improved upon the TPSC%
\cite{Tremblay98} developed long time ago to account for the spin and charge
fluctuations, which are the most important collective modes in the repulsive
Hubbard model. TPSC+ respects the spin and charge conservation laws, the
Mermin-Wagner theorem and the Pauli principle. It is based on the q and f
sum rules for the spin and charge susceptibilities. The spin and charge
susceptibilities, along with the double occupancy, are determined self
consistently with the TPSC+ ansatz, from which the self energy and Green's
function are evaluated.

The charge and the spin (Matsubara) susceptibilities, $\chi _{n\mathbf{p}%
}^{s,c}$, using the Bethe-Salpeter ansatz in the TPSC+ form (see details in
ref.\cite{Tremblay23}), are expressed in terms of the effective interaction
strength for the spin and the charge channels, $U_{s,c}$, and corresponding
susceptibilities: 
\begin{equation}
\chi _{n\mathbf{p}}^{s}=\frac{2\chi _{n\mathbf{p}}}{1-U_{s}\chi _{n\mathbf{p}%
}}\text{; }\chi _{n\mathbf{p}}^{c}=\frac{2\chi _{n\mathbf{p}}}{1+U_{c}\chi
_{n\mathbf{p}}}\text{.}  \label{eq1}
\end{equation}%
Here the second order correlator,

\begin{equation}
\chi _{n\mathbf{p}}=-\frac{2T}{N_{s}^{2}}\sum\nolimits_{m\mathbf{q}}\ G_{m%
\mathbf{q}}\left[ G_{m+n.\mathbf{q+p},}^{\left( 0\right) }+G_{m-n,\mathbf{q-p%
}}^{\left( 0\right) }\right] \text{,}  \label{eq2}
\end{equation}%
is expressed via free $G_{m\mathbf{q}}^{\left( 0\right) -1}=i\omega
_{m}-\varepsilon _{\mathbf{q}}-\mu \ $and the corrected , $G_{m\mathbf{q}%
}^{-1}=G_{m\mathbf{q}}^{\left( 0\right) -1}-\Sigma _{m\mathbf{q}}$, Greens
function. The self energy in turn (to the next to leading order order) is:

\begin{equation}
\Sigma _{m\mathbf{q}}=\frac{TU}{8N_{s}^{2}}\sum\nolimits_{m\mathbf{q}}\
\left( 3U_{s}\chi _{n\mathbf{p}}^{s}+U_{c}\chi _{n\mathbf{p}}^{c}\right)
G_{m+n.\mathbf{q+p}}^{\left( 0\right) }\text{.}  \label{eq3}
\end{equation}%
The set of equations Eq.(\ref{eq1},\ref{eq2},\ref{eq3}) for $\chi _{n\mathbf{%
p}}$, $U_{s}$ and $U_{c}$ was solved iteratively for sufficiently large
values of the sample size $N_{s}^{2}$ (periodic boundary conditions) with $%
N_{s}=128$ and sufficiently large number of Matsubara harmonics $\left\vert
m\right\vert \leqslant N_{t}=8192$ to access temperatures below $100K$.

The pairing energy is\cite{multilayerth}, 
\begin{equation}
V_{n\mathbf{p}}^{corr}=\frac{U}{2}\left( 3U_{s}\chi _{n\mathbf{p}%
}^{s}-U_{c}\chi _{n\mathbf{p}}^{c}\right) \text{,}  \label{Vcorr}
\end{equation}%
and was used to calculate the transition temperature along with the phonon
exchange. It is presented (for the IP layer of the tri-layer material,
phonon coupling $g_{a}=260meV$, $g_{p}=300meV$ and $U=3.5t$ ) at $n=0$ in
Fig. 6. One observes (see the left panel) a typical peak at the
crystallographic $M$ point. When the representation of the Brillouin zone of
this boson exchange is shifted by $\left( \frac{1}{2},\frac{1}{2}\right) 
\frac{\pi }{a}$ due to the antiferromagnetic origin of the paramagnon
intermediate boson. On the right panel of Fig. 6, a single $M$ point peak is
clearly seen.

\begin{figure}[h]
\centering\includegraphics[width=12cm]{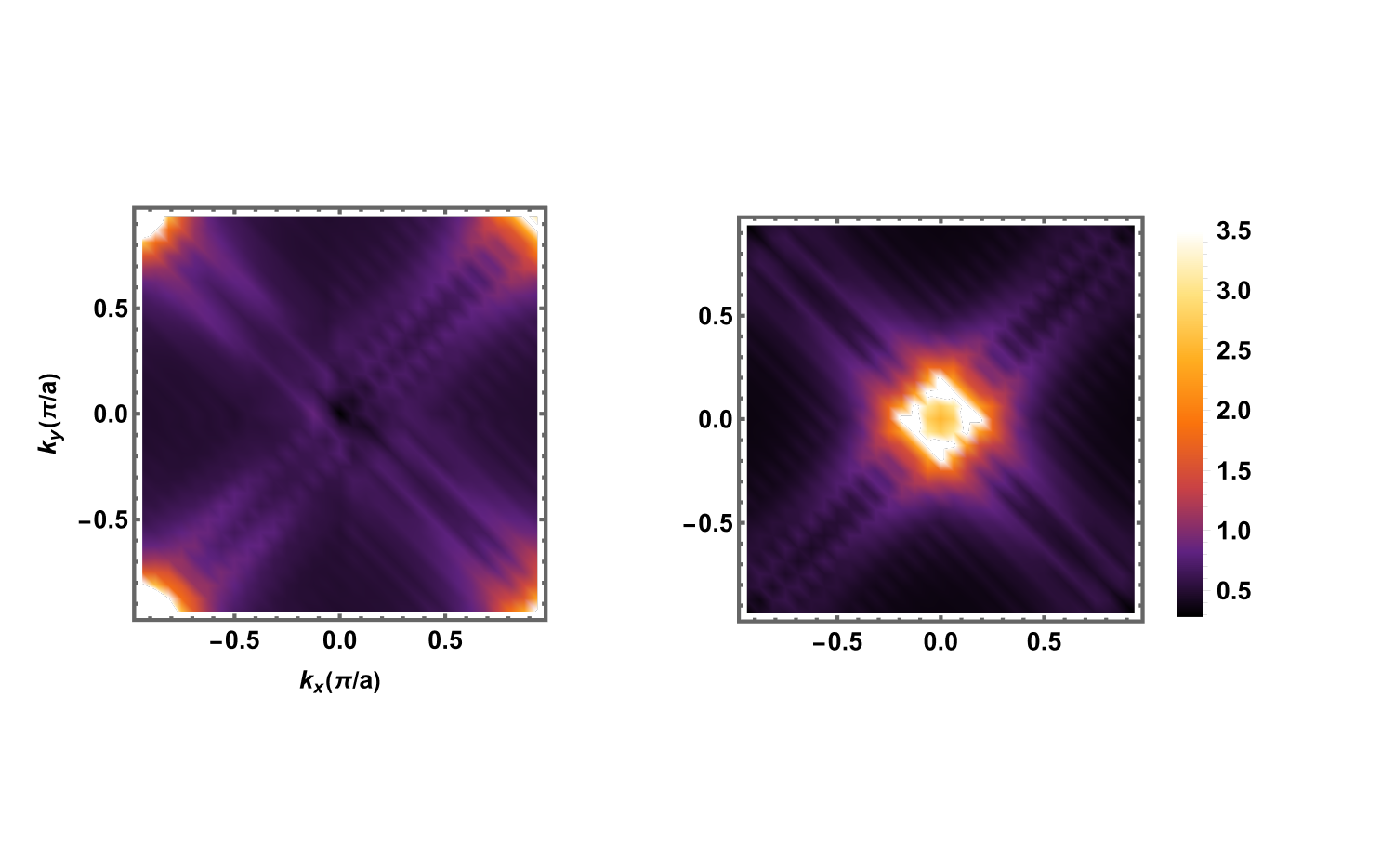}
\caption{ Spin fluctuation pairing potential. On the left the pairing
potential at zero Matsubara frequency calculated at $U/t=4$ is given. On \
the right the center of the boson's Brillouin zone is shifted to
crystallographic$M$ exhibiting a peak there.}
\label{fig6.pdf}
\end{figure}

When both terms of the effective interactions are added, one observes both
on the 3D plot (left panel) and the density plot (right panel) clear
separation of the gradients at the crystallographic $\Gamma $ point for
phonons and the $M$ point for SF. Only the gradients matter for the d-wave
pairing. It is different compared to the s-wave for which only the
homogeneous component is important.

\begin{figure}[h]
\centering\includegraphics[width=12cm]{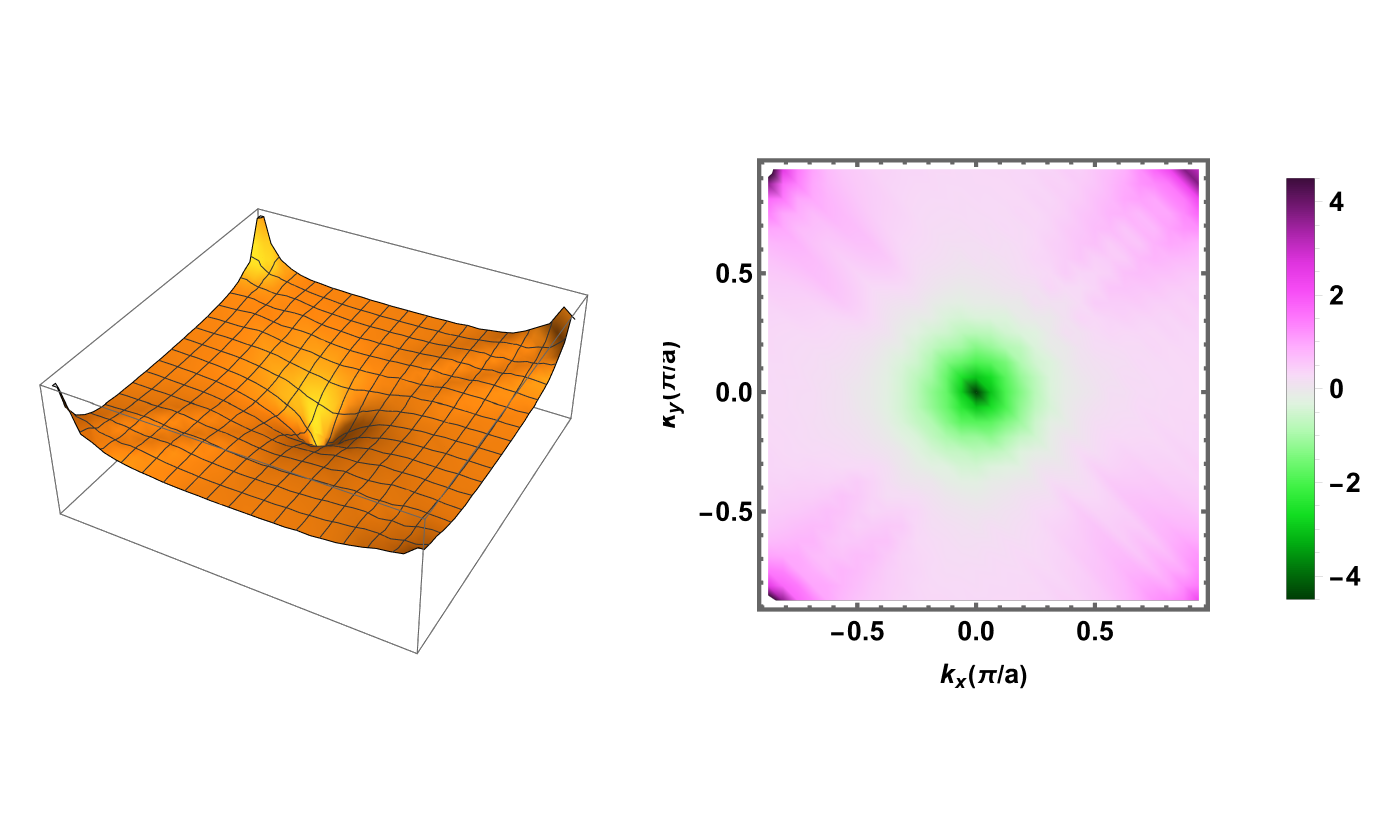}
\caption{ Phonon and spin fluctuations generating effective electron -
electron interaction are clearly separated. a. the 3D plot exhibiting the
phonon's central peak at crystallographic $\Gamma $ point and the SF with
peak at $M$ point is shown. b the density plot is provided to demonstrate
the separation.}
\label{fig7.pdf}
\end{figure}
This pairing potential is used in Section IV.

\section{The d-wave superconductivity in multi-layer materials.}

Transition to superconductivity is treated in a standard way since both
electron - electron effective interactions are not in the strong coupling
regime. This is of course consistent with the fact that $T_{c}$ is still
much lower than both interactions strength.

\subsection{Gap equations}

The standard superconducting gap equation at temperature $T$\ is,

\begin{equation}
\Delta _{m\mathbf{k}}=-T\sum\nolimits_{n\mathbf{p}}\frac{V_{m-n,\mathbf{k-p}%
}^{eff}\Delta _{n\mathbf{p}}}{\omega _{n}^{2}+\left( \epsilon _{\mathbf{p}%
}-\mu \right) ^{2}+\left\vert \Delta _{n\mathbf{p}}\right\vert ^{2}}\text{.}
\label{gapeqpara}
\end{equation}%
Here the (Matsubara) gap function is related to the anomalous Greens
functions $\left\langle c_{m\mathbf{k}}^{\sigma }c_{n\mathbf{p}}^{\rho
}\right\rangle =\delta _{n+m}\delta _{\mathbf{k+p}}\varepsilon ^{\sigma \rho
}F_{m\mathbf{k}}$ ($\varepsilon ^{\sigma \rho }$ - the antisymmetric
tensor), by 
\begin{equation}
\Delta _{m\mathbf{k}}=T\sum\nolimits_{n\mathbf{p}}V_{m-n,\mathbf{k-p}%
}^{eff}F_{n\mathbf{p}}\text{.}  \label{deltasupdef}
\end{equation}%
This is solved numerically. The gap equation was solved numerically by
iteration by discretization the BZ with $N_{s}=128$ and $32$ frequencies. It
converges to the d - wave solution, due to sufficiently strong direct
Coulomb part. Typically 500-1000 iterations are required (away from
criticality).

\begin{figure}[h]
\centering \includegraphics[width=12cm]{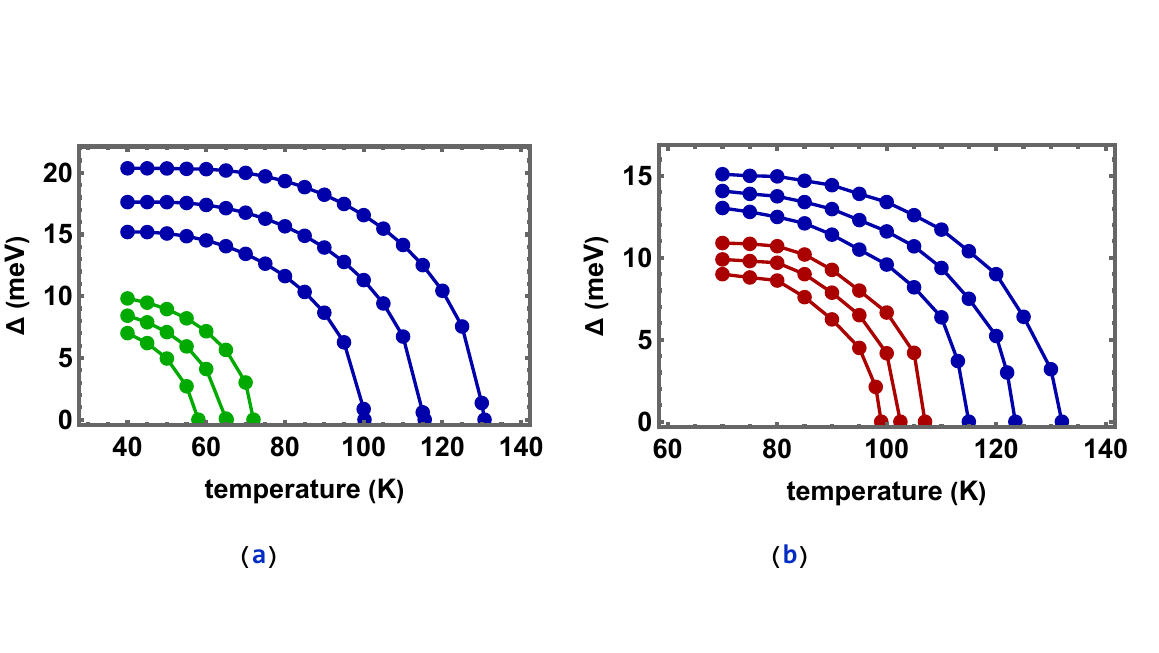}
\caption{Superconducting (Matsubara) gap as function of temperature. Two
sets of values of the electron - phonon couplings. Blue curves are for $%
g_{p}=300meV$, $g_{a}=250meV$ , while the green curves are for $g_{p}=200meV$%
, $g_{a}=155meV$. Number of layers are from one to three with the gap
increasing with the number of layers Values of the on-site repulsion are\ a. 
$U=3t$.\ b, $U=4t$.}
\label{Fig.8}
\end{figure}

\subsection{Results}

The results for $T_{c}^{\max }$ for values of $U=3t$ and$\ U=4t$ are given
for the single-layer, the bi-layer and tri-layered (IP) $Hg$ parameters in
Figs. 8a and 8b respectively. Values of optical phonon frequencies are kept
constant at $\Omega _{p}=80meV$ and $\Omega _{a}=55meV$. Same for the
spacings between the vibrating oxygen atoms and the conducting layers: $%
d_{p}=3.2A$, $d_{a}=2.7A$. Two different sets of the EPC are considered.
Blue lines correspond to $g_{p}=300meV$ and $g_{a}=250meV$, while green
lines are for $g_{p}=200meV$ and $g_{a}=155meV,$ The dependence on the
number of layers is indeed equidistant, but the "steps" are different.
Moreover it allows to conclude that $U=3t$, $g_{p}=300meV$ describes best
the experimentally observed values of $T_{c}^{\max }$ with differences of
about $15K$. The $U=4t$ smaller electron-phonon coupling becomes more
realistic, but then the differences are just about $10K$.

To demonstrate the constructive nature of cooperation between the phonon and
the SF exchange we present the calculation of the Matsubara gap function of
just phonons (no SF) or just SF (no phonons). If only the phonons are taken
into account, one obtains the dependence of the Matsubara gap $\Delta $ on
temperature shown in Figs.9a for three given values of EPC: $g_{p}=200meV$
(green), $250meV$ (blue)$\ $and $300$ $meV$ (brown). In Fig. 9b three values
of the gap for the three layers sample on-site repulsion values, $U/t=3$
(green), $3.5$ (blue) and$\ 4$ \ (brown), are calculated consistent with
Fig.1. One observes that both are much smaller than the combined gap shown
in Fig. 9.

\begin{figure}[h]
\centering \includegraphics[width=12cm]{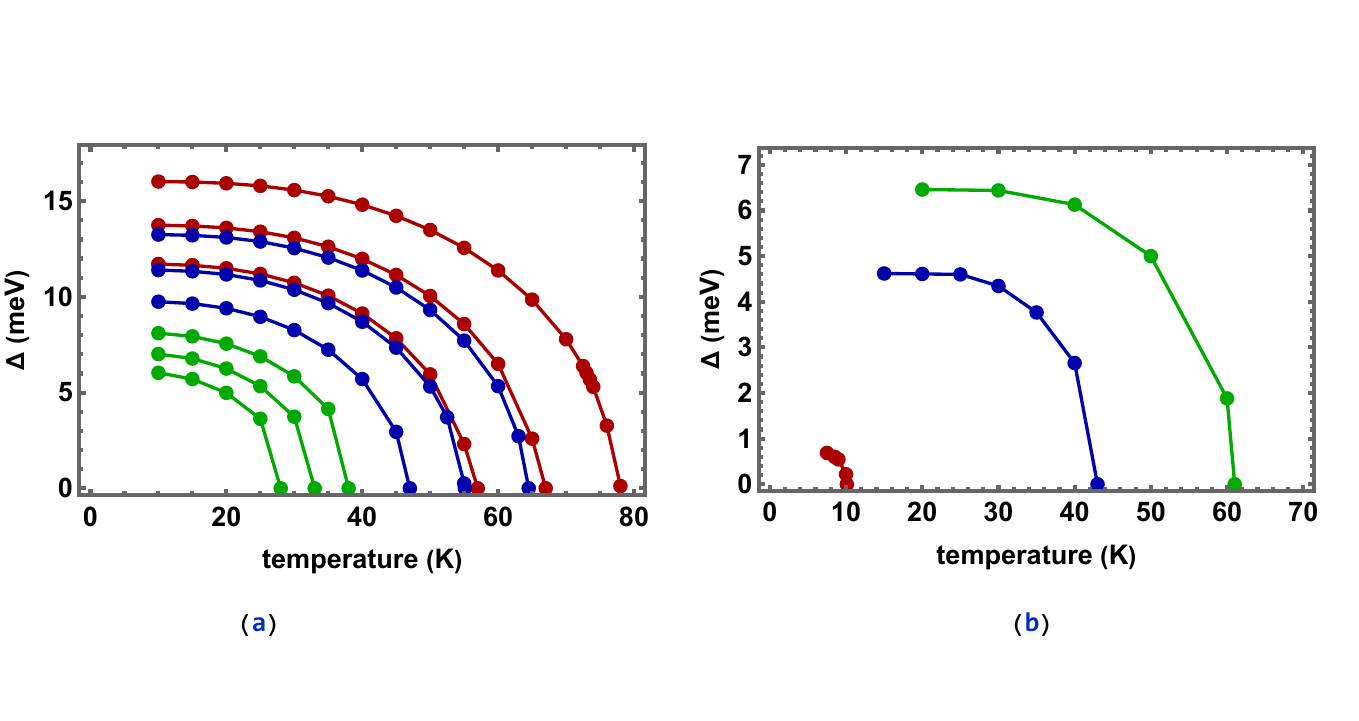}
\caption{Gap as function of temperature. \ Three curves in each groups). a.
phonons only (no magnons exchange) for three values of EPC $g_{p}=200meV$
(green)$,250meV$ (blue)$,300$ $meV$ (brown). For the same EPC the gap
increases with the number of layers (one, two and three are shown) b, spin
fluctuations only (no phonon exchange) for three given values $U=3t$ (green)$%
,U=3.5t$ (blue)$,U=4t$ (brown). Within the minimal model\ there is no
dependence on the number of layers.\ \ \ \ \ \ \ \ \ \ \ \ \ \ \ \ \ \ \ \ \
\ \ \ \ \ \ \ \ \ \ \ \ \ \ \ \ \ \ \ \ \ \ \ \ \ \ \ \ \ \ \ \ \ }
\label{Fig9.pdf}
\end{figure}

Let us now address the internal layer, IP, of the tri-layer material. While
in outer layers OP the dominant phonon modes are AP (on top) and FB and HB
(bottom), for IP one has FB and HB both on top and bottom, see Eq.(\ref{3L}%
). When we take the same chemical potential (means the same doping within
the minimal model) one obtains lower value of the critical temperature $%
T_{c}^{IP}=118K$ for the same parameters, namely critical doping for OP and
phonon and $U$ specified above. This might be consistent with experiment
that gives $T_{c}^{OP}=99K$ the doping of OP is lower that optimal and the
result should be even lower than calculated here\cite{largeN,Wen24}. Of
course the actual doping can be modeled by a more sophisticated model
describing the charging of inequivalent layers. This is clearly beyond the
minimal \ model and is briefly discussed in next Section.

\section{Discussion and conclusions}

To summarized, recently proposed\cite{I} apical oxygen atoms lateral
vibrations (AP) exchange mechanism of d-wave pairing in cuprates was
generalized to include a similar mechanism in multi-layer materials for
lateral $CuO_{2}$ planar oxygen vibration modes (essentially the full
breather, FB, and the half breather, HB). They generate the d-wave
superconductivity in the \textit{neighboring} $CuO_{2}$ layer (in addition
to the conventional\ in-plane s-wave pairing that is almost completely
destroyed by the Coulomb repulsion even for on-site repulsion as small as $%
U=2t$). Generally a lateral phonon mode in certain layer separated \ by the
spacing $d$ from the metallic $CuO_{2}$ layer generates attractive potential 
$V\left( k\right) \propto \exp \left[ -2kd\right] $. The d-wave pairing is
due to the peak of the potential at the crystallographic $\Gamma $ point of
(bosonic) Brillouin zone. In the present paper a "minimal" model describing
the electron gas in $CuO_{2}$ layers and phonons was used. Within this model
only two hoping terms are included, the nearest $t$ and the next to nearest
neighbor $t^{\prime }$ along with single interaction term, the on-site
repulsion $U$.

We calculate the critical temperature at optimal doping for a series $%
HgBa_{2}Ca_{N-1}Cu_{N}O_{2+2N}$ with number of layers $N=1,2,3$ with highest 
$T_{c}$. The model allows to explain the maximal critical temperature $%
T_{c}^{\max }$ dependence on the number of layers $N$. It rises by the steps
of $15K$ from $N=1$ to $N=3$ and then saturates. It is demonstrated that the
phonon exchange and the spin fluctuation pairing constructively enhance each
other. The strength of the on site Coulomb on site repulsion at optimal
doping is to obtain the observed values of $T_{c}^{\max }$ in the
intermediate range of $U=\left( 1.5-2\right) \ eV$, significantly smaller
than commonly considered in purely in-plane (spin fluctuation) theory of
high $T_{c}$ superconductivity. Such values however were obtained in some
first principle calculations\cite{DFTsmallU}. The electron - phonon coupling 
$g$ at the peak should be in the $\left( 200-300\right) meV$ range, also
within the values obtained in recent first principle studies\cite{Laoie21}%
\cite{Marzari23}. Qualitatively the explanation of the equidistant nature of
the critical temperature is as follows. The phonon pairing in a one-layer
material is just due to two AP modes (above and below the single $CuO_{2}$
layer), see Fig. 2 and Eq.(\ref{1L}). For a bi-layer one has three modes,
AP, FB and HB, see Fig. 3 and Eq. (\ref{2L}), while in the internal layer
(IP) of a tri-layer one has four modes participating, two FB and two HB, see
Fig. 4 and Eq.(\ref{3L}). As noticed above each phonon mode contribution is
about the same since $g_{AP}<g_{FB,HB}\simeq 1.3g_{AP}$ at $\Gamma $, is
balanced by the distances $d_{AP}=2,7A<d_{FB,HB}=2.3A$ effect. Interestingly
the outer layers (OP) of the tri-layer material seem to have lower gap\cite%
{Wen24} which is consistent with this argument, since they have only three
phonon modes pairing. The minimal model predicts that the OP $T_{c}^{\max }$
should be about the same as that of the bi-layer material. The model however
is insufficiently detailed to consider slow decrease of $T_{c}^{\max }$ at
higher $N$. In particular it was observed\cite{largeN} that at $N=6$ the
superconductivity at optimal doping becomes OP dominated.

Here we therefore add few remarks on the applicability range of the minimal
model and possible improvements. The most important terms left out are the
inter-layer hoping terms and the next to nearest neighbor screened Coulomb
repulsion term:%
\begin{equation}
\frac{W}{2}\dsum\limits_{\left\langle \mathbf{ij}\right\rangle }n_{\mathbf{i}%
}n_{\mathbf{j}}\text{.}  \label{Wterm}
\end{equation}%
The value obtained in recent \textit{ab initio} calculations\cite{DFText} is
about $W\approx U/5$. It might have a profound effect on the d-wave pairing.
Unlike the "large" on- site repulsion $U$ that does not influence the d-wave
pairing (while strongly suppressing the s-wave), this direct repulsion is
pair breaking in the d-channel. It actually plays a role of
"pseudo-potential" in the d-wave channel. Fortunately same calculations also
indicate that fitting the microscopic physics with a non-minimal model the
value of $U$ typically is larger (so perhaps the two effects balance each
other). The problem is that in this case the methods used here to treat the
SF in a relatively simple way are not available currently. Perhaps the
covariant GW method\cite{GW} can be adapted for this purpose. As far as $%
N\gg 2$ materials are concerned, one needs a microscopic charging theory
that determines dopings of inequivalent layers.

\bigskip

\textbf{Acknowledgements. }

We are grateful to D. Li, H.C. Kao and J.Y. Lin for helpful discussions.
Work of B.R. was supported by NSC of R.O.C. Grants No.
101-2112-M-009-014-MY3.\nolinebreak

\end{document}